# Participating or Not Participating? A Sociomaterial Perspective of the Embeddedness of Online Communities in Everyday Life

Geri Harris
School of Information Systems, Technology & Management
University of New South Wales
Kensington, Sydney
Email: geri.harris@unswalumni.com

Babak Abedin
Faculty of Engineering & IT
University of Technology Sydney
Sydney
Email: Babak.Abedin@uts.edu.au

## Abstract

Online communities are increasingly becoming a venue for socializing, engaging in politics, and conducting business. Ironically, the same enabling social–media technology is encroaching into everyday life and reconfiguring relations of participation. Yet, while participation in online communities has been widely studied empirically, theoretical aspects of this social phenomenon need further investigation. This paper uses a sociomaterial perspective to further develop theoretical explanation of participation in online communities and the impacts of *not* participating online. A sociomaterial view of online community participation decenters the human participant and recognises the agency of technology, thus creating a richer understanding than epistemological paradigms. Using converging hermeneutic circles, the paper first reviews literature for evidence of sociomaterial applications to online community research, and then proposes a framework for expressing participation in online communities from a sociomaterial perspective. Subsequently, implications of the findings and the potential for future studies are discussed.

## 1   Introduction

Social media is a topical issue drawing scholarly attention in refereed journals, reputable conferences and in the popular media (Kilpeläinen and Seppänen, 2014; Kwon et al., 2013; Niemi et al., 2013; Nyblom and Eriksson, 2014). However, technological advances enabling social-media based living are fast outstripping theoretical explanation, particularly in the understanding of how and why people participate online. The research problem motivating this study centres upon the potential for further theoretical understanding of participation in social-media based online communities and activities. Particularly in understanding the consequences of not participating online, for instance what happens to someone who is not a member of the online medium through which a social event is organized? Or what are the impacts of missing key moments in the lives of friends by not participating in their online community lives?

One might assume that the basic human desires for social, emotional, and physical support sought from membership of local communities, as claimed by Wellman and Wortley (1990), motivate people to interact in online communities. Online community theory (e.g. Barab et al., 2004; Preece and Maloney-Krichmar, 2005; Zhou, 2011) purports that these communities allow connections with wider networks; however, it reveals debate over the quality of electronically mediated interactions and the perceived value of relationships formed online. Online communities in everyday life are increasingly becoming a venue for socializing, engaging in politics, and conducting business (Gibson and Cantijoch, 2013; Halpern and Gibbs, 2013; Kumar and Singh, 2013; O'Murchu et al., 2004; Ravasan et al., 2014). Ironically, empirical evidence reveals that the same enabling social–media technology is increasingly encroaching into everyday life, reconfiguring the relations of online community participation as an assemblage of both the social and the technological (Harris, 2015; Boyd and Ellison, 2007; Qualman, 2012).

The significance of *not* participating in online communities, given that they are embedded in everyday life, centres – according to a recent study (cf. Harris, 2015) – upon what is missed out at both an individual and collective community level by not being online. Specifically, evidence reveals three areas in which effects of not participating are experienced by those outside social media–based online communities. These are,

(1)　　effects on relationships with friends, family, and community;

(2)　　limitations on participation in the life events of community members; and

(3)　　the inability to take advantage of community opportunities.

This paper expands current understanding of participation and nonparticipation in online communities by viewing the problem from an integrated research perspective. The sociomaterial worldview adopted recognises the embeddedness of online communities in everyday life in theorising important consequences of nonparticipation. The remainder of this paper is structured as follows: Firstly, the research problem and the significance of the study is presented in more detail (based on recent empirical findings). Section 3 presents a sociomaterial perspective to understanding online community participation. After that, the methodology by which this research was conducted is explained in section 4. Empirical sociomaterial research conducted in online community contexts is presented, analysed and discussed in sections 5 and 6. Finally, section 7 concludes by arguing that a sociomaterial research perspective assists in creating a different and richer understanding of online participation and nonparticipation than epistemological paradigms traditionally used in IS research.

## 2　Research Problem

### 2.1　Problem Domain and Research Objectives

The importance of participating in online communities and the impacts of not participating need to be better understood because – over time – membership of online communities becomes part of everyday life. A recent ethnographic field study (cf. Harris, 2015) illustrates this point, contributing important insights into participation and nonparticipation in online communities. Accounts from field study members (Harris, 2015) and claims at a societal level (e.g. DCITA, 2005) reveal that participating online is found to be motivated by individual needs for well-being, information sharing, autonomy, social contact, and entertainment. Further evidence reveals that online community participation materializes as changes to what is communicated and where/when it is communicated. Equally importantly, accounts from research participants reveal consequences of not participating in online communities. In the aforementioned field study (cf. Harris, 2015) members described their frustration when friends and family do not participate online, resulting in them missing out on life events that the field study members are communicating about in their online interactions with other friends and family members. It was also found that the relationships between members of the field study and their peers suffer when a member of the community with whom they have a relationship does not participate in community life online.

This current paper provides an exemplar of how a sociomaterial research perspective delivers a radically different and richer understanding of phenomena than epistemological paradigms traditionally used in IS research. Given the degree to which online communities are claimed to be embedded in everyday life in today's digital society (Kilpeläinen and Seppänen, 2014; Kwon et al., 2013; Niemi et al., 2013; Nyblom and Eriksson, 2014), it is difficult to understand participation by assuming a user existing separately from social media technology (having a separate existence of their own) (Barad, 2003, 2007; Orlikowski, 2000, 2007, 2010; Orlikowski and Scott, 2008a; Scott and Orlikowski, 2009, 2012). The problem domain becomes complex as social actors (users) and social media become intertwined, acting together in everyday practices of online community participation. Review of literature reveals existing research perspectives on participation in online communities treat the user (human) and technology (material) as ontologically separate (cf. Harris, 2015).

This paper argues that enquiry into participation and nonparticipation in online communities needs a fresh approach to understand the intertwining and mutual co-creation of the human/social and the technological. Specifically, the objectives of this paper are to,

(i)　　articulate the problem of treating the human and technological as separate and discreet aspects of online community participation;

(ii)　　critically analyse exemplary empirical studies that have taken a sociomaterial approach to uncovering new and unique insights about elements of online community participation that are possible from such a research perspective; and

(iii)　　present and explain the contributions that can be added to the current literature when an integrated research method is used to treat the human and technological aspects of online community participation as inseparable.

### 2.2 Research Significance

The importance of this paper is that it challenges long-held assumptions about how humans appropriate technology to communicate with one another. It progresses a theoretical understanding of participation in online communities in general, and in social media in particular, that accommodates the view of the human and the technological aspects as intrinsically inseparable in everyday instances of online community participation. Analysis also demonstrates that what *is* known is limited by an ontology of separation between the social and the technological (Barad, 2003; Orlikowski, 2007; Kautz and Jensen, 2012; Cecez-Kecmanovic et al., 2014a). The key issues arising from privileging either a technologically deterministic perspective of participation or an equally autonomous social-centric perspective include (Harris, 2015):

- Understanding of the effects of ICT in social life and participation in online communities is limited by the dualistic treatment of the social and technological aspects as separate entities existing independently of one another.
- Current understanding of the role of technology in social life centres on the changes it makes to the nature of interpersonal relationships and the resulting creation of altered forms of and spaces for social contact. Existing theory focuses on understanding the technological aspects affecting social life, with much less evidence uncovered of theorizing technology in social life as a social phenomenon.
- A picture of a different community that forms and exists online emerges from literature analysis. The concept of community is found to change when it is enacted online, with evidence supporting this changing nature of community as a direct effect of the proliferation of ICT in everyday society. Review of literature exposes an emphasis within existing theory on the social aspects of technology in social life.
- Changes to the concept of socialising in the digital age affect the conceptualisation of community, discussed primarily from a human-centric perspective that community changes when it is enacted online.
- Existing theory on the phenomenon of participation discusses the difficulty in defining it for its multiple meanings in different contexts and broad categorisations of the levels at which participation occurs
- A human-centric view posits that social inclusion and exclusion manifest differently in online communities.

The rationale for this paper is to show that there is evidence in literature that social and technological aspects of phenomena overlap and work interdependently to become sociomaterial (cf. Barad, 2003; Cetina, 1997; Latour, 2005; Orlikowski, 2009; Pickering, 1995; Schatski, 2002) over time. Analysis of exemplary empirical studies that examine elements of online community demonstrates that everyday participatory practices are shown as performed within assemblages. Specifically, sociomateriality (Barad, 2003, 2007; Orlikowski, 2000, 2007, 2010; Orlikowski and Scott, 2008a; Scott and Orlikowski, 2009, 2012) is put forward as an appropriate perspective from which to explain participation and nonparticipation in online communities and activities because of its integrated perspective of treating the human and nonhuman as inseparable (cf. Cecez-Kecmanovic et al., 2014a; Kautz and Jensen, 2012).

## 3 Sociomateriality: A Fresh Perspective on Online Community Participation

A sociomaterial interpretation of social phenomena like participating in online communities establishes a unique position for its view of people intersecting with technology to participate in digitally-enabled social networks (Cecez-Kecmanovic et al., 2014a, 2014b; Doolin and McLeod, 2012; Orlikowski and Scott, 2013; Scott and Orlikowski, 2012). The philosophical assumptions underlying sociomaterial research position sociomateriality as a post-humanist research perspective. It moves beyond a socio-technical position by decentering the 'human' subject and recognising the agency of the 'nonhuman' in participating in online communities (Gherardi, 2009; Latour, 2005; Orlikowski, 2000). This creates the opportunity to uncover a rich tapestry of relational understanding of online community participation that does not privilege either the social or the material. In such a non-essentialist worldview, technology, as a nonhuman with agency, has the ability to be something else other than the essential properties it is designed with and given. This perspective "makes evident the importance of taking account of 'human,' 'nonhuman,' and 'cyborgian' forms of agency" (Barad, 2003 p. 826) and "does not fix the boundary between 'human' and 'nonhuman'" (Barad, 2003 p. 821). The key ontological difference is a debate over separation versus relationality. Criticised as dualist ontological perspectives, traditional established research approaches based upon a Cartesian worldview assume the separate existence of entities with defined a priori attributes; the world is viewed as something external and we make representations of this.

Several sources of sociomaterial concepts exist, for example in the works of Barad, 2003, 2007; Orlikowski, 2007, 2010; Orlikowski and Scott, 2008; Suchman, 2007; and Leonardi, 2013. Orlikowski posits that a sociomaterial view of the world is premised on the notions of *constitutive entanglement*, *sociomaterial assemblages*, *performativity*, *intra-action*, and *temporal emergence*. These can be considered second-order concepts, all founded on a philosophical position of relationality, concepts that can be applied methodologically in conducting sociomaterial research. The IS community positions sociomateriality as a "new lens" for research that "questions the givenness of the differential categories of 'human' and 'nonhuman'" (Kautz and Jensen, 2012 p. 808). Sociomateriality is an emerging research perspective, however it is suitable for studies of social phenomena because of its ability to bring together the social and the technological aspects better than its predecessors. In particular, sociomaterial practices are suited to being applied as a theoretical lens in making sense of data gathered from inquiry into online community participation/nonparticipation, and as a framework within which to interpret empirical findings in sociomaterial terms. Framing investigation of the participation and nonparticipation of particular individuals in online communities and activities as sociomaterial research allows enquiry into the reconfiguration of participatory practices when online community participation is possible. Sociomateriality, as demonstrated in upcoming sections 5 and 6, has the ability to bring the social and the technological aspects of online community participation together in a way that considers the human and the nonhuman to be intertwined in constructing everyday reality.

## 4   Research Method

The methodology used to conduct the research presented in this paper was a two-stage process of developing an understanding of relevant literature through hermeneutic circles (Boell and Cecez-Kecmanovic, 2014). The first stage involved conducting a methodological search of IS literature to locate and acquire relevant papers that have approached enquiry into elements of online communities, social media and the phenomenon of participation from a sociomaterial perspective. The second stage involved mapping and classifying acquired literature identified as contributing a sociomaterial research perspective of these elements of online communities and participation as social phenomena. This literature was critically assessed in developing the argument for further explanation of participation /nonparticipation in in an integrative manner. Both stages of seeking and critically assessing information are intertwined and converge to produce a "well-grounded, novel and interesting outcome" (Boell and Cecez-Kecmanovic, 2014 p.264) in the problematization of understanding online participation from existing separatist research perspectives.

### 4.1   Literature Searching and Acquisition

The literature search process took the form of a hermeneutic circle of searching, sorting and filtering existing literature in order to locate material of specific relevance to the research problem and objectives articulated in section 2. A circular process of literature search and acquisition was conducted to ascertain what is currently known about sociomateriality as a research method and its application to investigating elements of participation and online communities as social phenomena. Adapted from the work of Boell et al. (2014), Figure 1 depicts the circular approach to searching, sorting, selecting, acquiring, reading, identifying and refining the literature review process in a hermeneutic circle. Concentrated within the IS and social science disciplines, literature searches targeted high-ranking journals, including MIS Quarterly, Information Systems Journal (ISJ), Journal for the Association of Information Systems (JAIS), Organization Science, Organization, and Communications of the ACM. Literature analysis moved from searching individual literature sources on sociomateriality, including refereed journal articles and textbooks, to the wider field of knowledge on research methods and back again several times in arriving at conclusions. Searches were conducted within databases including EBSCO, ScienceDirect, Scopus, and Google Scholar.

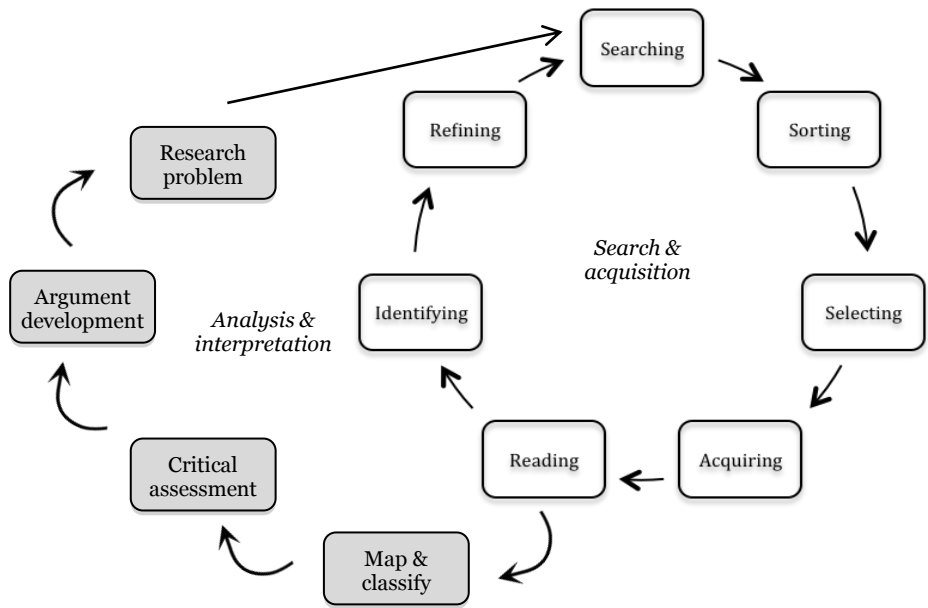

*Figure 1. Hermeneutic cycle approach to literature review process*

### 4.2 Assessment and Interpretation of Literature

Subsequent to the identification of empirical examples of research applying elements of sociomateriality to the enquiry of online communities and/or participation, a second and wider hermeneutic circle of analysis and interpretation was conducted. In an iterative manner literature was read to increase understanding of sociomateriality and its application to empirical problems in the specific examples acquired during the search and acquisition phase. Adding to sociomateriality the central terms of online community, participation and nonparticipation, the literature circle continued by identifying the main authors and core high-ranking journals containing relevant material. Literature was further mapped and classified through successive fractions and through an iterative process arrived at a selection of empirical studies that had relevance to some elements of participation in online community. Database searches located a body of theory pertaining to sociomateriality originating from a few key authors and a relatively small number of empirical examples of sociomaterial research. Cited publications were reviewed and annotated in EndNote[1], extrapolating central sociomaterial concepts and terms, main authors, and exemplary studies. This process was repeated until the main literature had been covered, as gauged by confidence that the well-cited publications and key authors of sociomaterial research had been examined and there was "high confidence in the novelty and importance of a contribution" (Boell and Cecez-Kecmanovic, 2014, p.272) to the problem of understanding online community participation and nonparticipation.

An alternative to the systematic literature review (Kitchenham, 2004), a hermeneutic approach to understanding and problematizing participation and nonparticipation facilitated a dynamic, iterative process of literature review, transparent analysis and critique (Boell and Cecez-Kecmanovic (2014). Results from this analysis of existing knowledge conclude that within IS research, literature of direct relevance to sociomateriality and its application as a research method to empirical studies is limited. Analysis reveals that sociomateriality is an emerging research perspective and methodology that offers the possibility of adopting a fresh approach to conducting IS research. Building upon the conclusions drawn from analysis and critique of literature, the next step in the research was to develop an evidence-based argument for adopting a sociomaterial research perspective to advance understanding of online community participation and nonparticipation.

## 5　Developing the Argument from Evidence in Literature

This section builds from the earlier stages of classification and analysis of relevant literature. The findings are presented from critical assessment and interpretation of empirical research within the IS discipline in which sociomaterial concepts have been applied. Such analysis involved the identification and substantive description of sociomaterial concepts which were then applied to illuminate what particular studies found that is sociomaterial. Subsequently in describing how this worldview helps to understand IS and social phenomena in a different way, analysis of these studies illustrates that "focusing on sociomaterial aspects of everyday practices will open up important avenues for examining and understanding ongoing production of [social] life" (Orlikowski, 2007 p. 1445). Reflecting on the sociomaterial understanding of the phenomena in these studies now provides a solid foundation upon which to develop an argument for approaching enquiry into online community participation/nonparticipation from an understudied, integrative perspective.

---

[1] EndNote is a software tool for publishing and managing bibliographies, citations and references.

Table 1 summarises empirical examples of sociomaterial studies that are relevant to a theoretical understanding of online communities and participation. Whilst literature searching did locate a wider body of empirical sociomaterial research[2], only studies with direct relevance to some aspect of participation/nonparticipation have been analysed for their contribution to a sociomaterial perspective of information systems. These studies illustrate that "focusing on sociomaterial aspects of everyday practices will open up important avenues for examining and understanding ongoing production of [social] life" (Orlikowski, 2007 p. 1445). Collectively, the findings illustrate that sociomateriality offers novel and interesting insights into IS and social phenomena where dualist research perspectives cannot. Furthermore, evidence reveals the embeddedness of technology in everyday organisational and social life. For each study, the phenomenon of interest is highlighted together with a summary of the key sociomaterial practices that are enacted.

| Empirical example | Phenomenon: | Summary of sociomaterial perspective |
|---|---|---|
| TripAdvisor ratings<br>Scott and Orlikowski (2012) | Hotel travel practices: accountability online | In the performance of online ranking, social media is entangled in everyday practices of hoteliers and travellers. The TripAdvisor website is integrated into the practices of travellers planning travel arrangements online. Illuminates reconfiguring relations of accountability, wherein "accountability is always and unavoidably an inseparable, sociomaterial entanglement" (p. 36). |
| TripAdvisor and the Automobile Association (AA)<br>Scott and Orlikowski (2014) | Hotel travel practices: anonymity | Anonymity online is a dynamic material enactment, constituted in practice through ongoing materialisations. Highlights "the different line of inquiry that sociomateriality inspires and how it reframes issues…that would otherwise presume separate entities" (p. 3). |
| TripAdvisor<br>Orlikowski and Scott (2014) | Hotel travel practices: evaluations online | Online travel hotel evaluation is a result of considering the performance of online valuations within hotel rating practices situated within a sociomaterial assemblage. Travel is "performed differently now that algorithmic valuation apparatuses such as TripAdvisor exist" (p. 887). |
| The algorithm and the crowd—the materiality of service innovation<br>Orlikowski & Scott (2015) | Hotel travel practices: service innovation | Service innovations are material-discursive practices performed in emerging crowd-sourced algorithmic transformations. Emphasises the relationality and materiality in contemporary online service innovation, focusing on understanding how "boundaries are drawn…phenomena are configured, and what realities are performed" (p. 14). |
| Mobile communications at Plymouth organisation<br>Orlikowski (2007) | Mobile communication | The "performativity of BlackBerry's as engaged in members' everyday practices" is sociomaterial (p. 1444). Communication is entangled with BlackBerrys, resulting in a "blurring of employees' work and personal lives" (p. 1444). |
| Participation in Online Communities | Online parenting | Participation is enacted online and offline contemporaneously through entanglements of social |

---

[2] Literature searches located 12 empirical sociomaterial studies from IS literature.

| | | |
|---|---|---|
| Harris (2015) | community participation | actors, social media, community values, beliefs, norms and rules for communication. Illustrates the intra-actions in everyday instances of participating in community life on- and offline. |
| Olympia online project IS assessment<br><br>Cecez-Kecmanovic et al. (2014) | IS assessment | IS success and failure are "performed and thus determined by sociomaterial practices" (p. 561) in IS–project actor networks. Success or failure is "the enactment of an information system in sociomaterial practices emerging through specific intra-actions among actors" (p.567). |
| Knowledge workers' time management<br><br>Kahrau et al. (2013) | Time management | A performative account of practices in (1) remembering tasks, (2) deciding what to do next, and (3) maintaining a well-organised workplace. |

*Table 1. Application of sociomaterial practices in existing empirical studies*

Analysis of these studies illuminates the performance of everyday organisational and social practices as enacted within sociomaterial assemblages. For example, Scott and Orlikowski (2012) in their TripAdvisor study exemplify how the performance of online valuations within hotel rating practices is situated within a sociomaterial assemblage. For TripAdvisor, a sociomaterial assemblage is constituted as hotel reviews, TripAdvisor members, TripAdvisor website, computer, the Internet, browser software, sign-in procedure, review writing activities, on-screen feedback, representation of hotels, other reviews, databases, rating and ranking mechanisms, verification protocols, and e-mail communication. For AA, a sociomaterial assemblage is constituted as hotel, inspector, inspection activities, inspector training/experience, knowledge of standards, engagement with quality criteria, spreadsheets, observations, recordings, reports, discussions with hotel staff, editors, and other inspectors. Table 1 demonstrates that when a sociomaterial perspective is applied to empirical research, unique insights are revealed in the understanding of IS and social phenomena. Online activities such as hotel ranking, IS assessment, anonymity, travel evaluations, service innovation, communication and participation are found to be both performed and enacted in assemblages that are an integration of the social and the technological. Extending these findings to understanding instances of participation in online communities and activities reconfigures communication and information sharing and how these make some practices more salient than others. Reflecting on the sociomaterial understanding of the phenomena in the studies identified provides an analytical framework within which to extend current understanding of online community participation/nonparticipation. Section 6 further discusses the application of a sociomaterial perspective and its core concepts to understanding participation online.

## 6    Analysis of Findings

The unique perspective of this paper is to look at the problem of limited understanding of participation and nonparticipation in online communities from a sociomaterial perspective. In table 2 key concepts of sociomateriality (cf. Barad, 2007; Orlikowski, 2007; Orlikowski, 2009) are applied to the problem domain. Expressing aspects of online participation as sociomaterial practices allows this paper to demonstrate the sociomaterial accomplishment emerging from enmeshment of social actors and context with enabling media technology.

| Practice | Definition |
|---|---|
| Relationality | Online communities exist in relation to other assemblages. That is, within an online community, common interest unites agents, whereas across communities, differences in practices (such as participant needs) will create boundaries and potential conflict. |
| Performativity | Relationships between community participants and social media technology are never fixed. Instead, the sociomaterial assemblage (online community) emerges from practice and defines how to practice. It is in the act of participating that the relation (between the participant and the social media technology) is defined, and each participatory act produces (or performs) a different relationship. |
| Entanglement | The material and the social emergently produce one another, as people, entangled with a variety of social media–based technologies, participate in online communities in the carrying out of their daily social practices. |
| Sociomaterial assemblage | An online community is a composite and shifting assemblage of the material and social, which change over time as those involved participate in online community activities to provide meaning, to exercise power, and to legitimate actions. |
| Co-constitution | The material (media technology) and the social (community participants) are mutually constituted and inseparable. Structures and processes of an online community are enacted and emergent as participants draw upon communication features in their situated practices. |

*Table 2. Online community participation expressed in sociomaterial terminology*

Inquiry into participation and nonparticipation in online communities requires a theoretical perspective that recognises technology and everyday practices as intrinsically linked. Technology and human agency cannot ontologically be studied separately because they are always in relation; therefore, the social (human) and the material (technological) aspects must be treated as inextricably linked, a notion proposed by Barad (2003). Viewing the social and material aspects of participation as intertwined in the construction of everyday reality, IS assumptions of separateness are challenged. A sociomaterial view holds that aspects of online community participation are entangled in everyday social practices and cannot be understood in isolation. Research can then examine the mutual co-construction of participation/nonparticipation to understand how meanings and materialities are enacted in everyday social and technological practices (Barad, 2007; Introna, 2007; Suchman, 2009).

Drawing on the exemplar empirical papers in Table 2 that adopt and develop a sociomaterial approach (e.g. Orlikowski and Scott, 2012, 2013), the utilisation of sociomateriality as a theoretical lens furthers scholarly understanding of online community participation and nonparticipation as sociomaterial practices in several ways. First, because established, competing perspectives on human-technology research are problematic in privileging either the technological or the human/social factors, sociomateriality gives agency to the nonhuman (social) and treats the human and the social as always in relation, mutually co-constructing each other. Second, because sociomateriality allows for the contextualisation of participation in a sociomaterial setting, it creates the possibility for perceiving both the technological and social contexts in a more integrated way (Orlikowski, 2009). Third, from a sociomaterial perspective, participation and nonparticipation are framed as enacted in material-discursive practices where, as described by Iedema (2007), social media technologies have no inherent boundaries or meaning but are bound up with specific material-discursive practices—material-discursive practices that constitute online community participation.

Participation and nonparticipation in online communities are concluded to be performed as enmeshments of social actors and context with enabling social media technologies. Based on the empirical examples reviewed, the problem of participation and nonparticipation means the questioning of the taken-for-granted, essentialist nature of entities. In this sense, online community participation is emergent in nature and enacted in everyday instances of participating, within which participants, nonparticipants and the technology are "temporally constituted by discursive-material practices" (Cecez-Kecmanovic et al., 2014a p. 566). In sociomaterial terminology online community participation and nonparticipation are temporal and emergent. Theoretically the role of technology in everyday online activities becomes such that technology is materialised beyond a question of adoption versus non-adoption. This richer understanding recognizes the intertwined nature of communicative actions and social media when participating or choosing not to participate online.

## 7  Conclusions and Agenda for Future Research

Based on critical analysis of a wide spectrum of IS and social theory (cf. Harris 2015) and the analysis in this paper of empirical studies, a sociomaterial research perspective assists in creating a different and richer understanding of phenomena such as online participation and nonparticipation than epistemological paradigms traditionally used in IS research. This perspective challenges established, competing perspectives that privilege either a technological or a human-centric understanding of social phenomena. Nonparticipation and participation (and anything in between) are part of an emergent holistic phenomenon of living a life entangled in complex social and technological environments. The intra-actions going on among the entangled actors produce diverse effects - from highly valuing to not valuing online participation; from enjoying to disliking online participation; from being online all the time to avoiding online communication. Apart from these effects, the intra-actions are also (re)producing the community and what it means to belong to a community. All actors are in a continuous process of change suggesting that intensity of online participation (including nonparticipation) is also changing. It does not seem fruitful to seek explanation by identifying factors that determine participation or non-participation. Rather, it looks more promising to increase understanding of the phenomenon of participation as a holistic phenomenon emerging in a complex social and technological environment.

This paper opens new opportunities for future research into online and social communities and technologies. One opportunity for a deeper understanding is to recognise the agency of social media intertwined with individuals in enacting the practice of participating in community online. To understand what it means to socialise in a digital age is to understand what it means when an individual participates (or does not participate) in the life and activities of their community by engaging in face-to-face interactions, online interactions or both simultaneously by being present in one physical setting and potentially mentally engaged with something external to this physical setting. Similar research can be done to explore different types of communities, development of professional and private relations in these communities, as well as the implications felt by participating and nonparticipating members. Further research is needed to examine other aspects of online community participation and nonparticipation, including how and where norms and values are defined and negotiated as part of community emergence and continuing transformation and how members of online communities become socialized into these norms within which they are expected by society to interact. Future studies could investigate the impacts and/or implications of participation versus nonparticipation in specialized online communities, such as health or education communities, from a sociomaterial perspective. Since more people are using specialized online communities for information and experience exchange for purposes such as psychological wellbeing (Erfani et al., 2013), healthcare (Bender et al., 2011), and learning and personal development (Abedin, 2012), it is becoming necessary to look at human-computer interactions from a fresh sociomaterial approach. Such an approach to empirical enquiry within a context of these complex online social communities creates the possibility for research that is innovative in advancing understanding of participatory practices and the impacts for nonparticipating individuals.